\documentclass[a4paper]{jpconf}
\usepackage{graphicx}
\usepackage{amsmath}

\begin{document}
\title{Quantum information aspects of noncommutative quantum mechanics}

\author{Orfeu Bertolami, Alex E. Bernardini and Pedro Leal}
\address{Departamento de F\'isica e Astronomia, Faculdade de Ci\^encias da Universidade do Porto, Rua do Campo Alegre, 687,4169-007 Porto, Portugal}
\ead{orfeu.bertolami@fc.up.pt}
\begin{abstract}
Some fundamental aspects related with the construction of Robertson-Schr\"odinger like uncertainty principle inequalities are reported in order to provide an overall description of quantumness, separability and nonlocality of quantum systems in the noncommutative phase-space. Some consequences of the deformed noncommutative algebra are also considered in physical systems of interest.
\end{abstract}

\subsubsection*{Introduction --}
In searching for a broader understanding of the transition from quantum to classical description, noncommutative (NC) algebra deformations of the standard quantum mechanics (QM) plays a relevant role. The NC extension of QM is characterized by the deformation of the Heisenberg-Weyl algebra,
\begin{equation}
[\hat{x}_i, \hat{x}_j] = i\theta_{ij}, \quad [\hat{x}_i, \hat{p}_j] = i\hbar \delta_{ij}, \quad [\hat{p}_i, \hat{p}_j] = i\eta_{ij}, \quad i = 1,...n,
\label{ggreqs01}
\end{equation}
where $\theta_{ij}$ and $\eta_{ij}$ are skew-symmetric matrices with real entries, whereas in the standard quantum mechanics one has
\begin{equation}
[\hat{q}_i, \hat{q}_j] = 0, \quad [\hat{q}_i, \hat{k}_j] = i\hbar \delta_{ij}, \quad [\hat{k}_i, \hat{k}_j] = 0, \quad i = 1,...n,
\label{ggreqs02}
\end{equation} 
with the two algebras being connected by the so-called Seiberg-Witten (SW) map \cite{Seiberg},
\begin{equation}
\hat{x}_i = \nu\hat{q}_i -\frac{\theta}{2\nu\hbar}\epsilon_{ij} \hat{k}_j, \quad \hat{p}_i = \mu \hat{k}_i + \frac{\eta}{2\mu\hbar}\epsilon_{ij}\hat{q}_j,
\label{ggreqs03}
\end{equation}
which allows, through constants $\nu$ and $\mu$, mapping NC effects in terms of standard QM variables.

From the perspective of a deformed Heisenberg-Weyl algebra of QM \cite{Bastos,Gamboa,Rosenbaum}, several issues on NC QM related to missing information in gaussian quantum systems \cite{Bernardini13B,Bernardini13B2,Salomon}, quantum correlations and information collapse \cite{Bernardini13A,2015,2016}, and violations of the uncertainty relations \cite{Bastos3,RSUP1,RSUP2,RSUP3} have been investigated.

The theoretical framework of the NC QM \cite{Bastos,Gamboa,06A,08A,09A} is suitably formulated in the Weyl-Wigner-Groenewold-Moyal (WWGM) formalism for QM \cite{Groenewold,Moyal,Wigner}, in a theory supported by a $2n$-dim phase-space where coordinate and momentum variables obey a NC algebra.
Through the  WWGM formalism, the NC QM has been used to understand the coupling and decoherence aspects of $2$-dim quantum oscillators \cite{Rosenbaum,Bernardini13A}, the quantum Hall effect \cite{Prange} and the Landau level problem \cite{Nekrasov}, corrections to the gravitational quantum well \cite{06A} and nuclear matter \cite{Mariji_2015}.
Furthermore, besides its relevance in the context of quantum gravity and string theory \cite{Seiberg,Connes,Douglas}, NC extensions of quantum cosmology have also been considered in the investigation of the black hole singularity \cite{Bastos3,Bastos3B,PedroLeal}, as well as of gauge invariance \cite{Queiroz_2011,Leal} and the equivalence principle \cite{Leal,Bastos4}.

\subsubsection*{Quantumness --}
Considering a bipartite quantum system described in terms of a $2n_A$-dimensional subsystem $A$ and a $2n_B$-dimensional subsystem $B$, with $n_A+n_B=n$, one can write the collective degrees of freedom of the composite quantum system as
$\widehat{z} = (\widehat{z}^A,\,\widehat{z}^B)$, where $\widehat{z}^{A,B} =(\widehat{x}_1^{A,B},
\cdots,\widehat{x}_{n_{A,B}}^{A,B},\, \widehat{p}_1^{A,B}, \cdots, \widehat{p}_{n_{A,B}}^{A,B})$ correspond to the generalized coordinates of the two subsystems \cite{Bernardini13B}.
The NC phase space algebra given by 
\begin{equation}
\left[\widehat{z}_i, \widehat{z}_j \right] = i\, \Omega_{ij}, \hspace{1 cm} i,j = 1, \cdots, 2n,
\label{ggreqs1}
\end{equation}
is described by the matrix, ${\bf \Omega} = \left[\Omega_{ij} \right] \equiv \mathrm{Diag}\left[\bf \Omega^A,\,\bf \Omega^B\right]$, where \cite{Bastos,Bernardini13B}
\begin{equation}
{\bf \Omega^{A,B}} = \left(
\begin{array}{r r}
{\bf \Theta^{A,B}} & {\bf I^{A,B}}\\
- {\bf I^{A,B}} & {\bf \Pi^{A,B}}
\end{array}
\right),
\label{ggreqs2}
\end{equation}
are real skew-symmetric non-singular $2n_{A,B} \times 2 n_{A,B}$ matrices with ${\bf \Theta^{A,B}} = \left[\theta_{ij}^{A,B} \right]$ and ${\bf \Pi^{A,B}} = \left[\eta_{ij}^{A,B} \right]$ measuring the noncommutativity of the position and momentum sectors of the subsystems $A$ and $B$, respectively\footnote{Here ${\bf I^{A,B}}$ is the $n_{A,B} \times n_{A,B}$ identity matrix and it has been set $\hbar =1$.}.
The generalization of the NC structure from Eqs.~(\ref{ggreqs01})-(\ref{ggreqs03}) is indeed formulated in terms of commuting variables through the linear Darboux transformation (DT) (also referred to as SW map), Eq.~(\ref{ggreqs03}), $\widehat{z} = {\bf S} \widehat{\zeta}$, with ${\bf S} \in \mathrm{GL}(2n)$ such
that ${\bf S} = \mathrm{Diag}\left[{\bf S^A},\,{\bf S^B}\right]$, where $ \widehat{\zeta}=( \widehat{\zeta}^A,
\widehat{\zeta}^B)$, with $ \widehat{\zeta}^{A,B}= ( \widehat{q}_1^{A,B}, \cdots, \widehat{q}_{n_{A,B}}^{A,B}, \, \widehat{k}_1^{A,B},
\cdots, \widehat{k}_{n_{A,B}}^{A,B})$ satisfying the usual QM commutation relations:
\begin{equation}
\left[\widehat{\zeta}_i, \widehat{\zeta}_j \right] = i J_{ij} , \hspace{1 cm} i,j=1, \cdots, 2n,
\label{ggreqs3}
\end{equation}
where ${\bf J} = \left[J_{ij} \right]=\mathrm{Diag} \left[{\bf J^A},{\bf J^B} \right] $ with
\begin{equation}
{\bf J^{A,B}} = - ({\bf J^{A,B}})^T = - ({\bf J^{A,B}})^{-1} = \left(
\begin{array}{r r}
{\bf 0^{A,B}} & {\bf I^{A,B}}\\
- {\bf I^{A,B}} & {\bf 0^{A,B}}
\end{array}
\right)~,
\label{ggreqs4}
\end{equation}
which are $2n_{A,B} \times 2n_{A,B}$ standard symplectic matrices.
Likewise the SW map, also the map ${\bf S}$ is not uniquely defined: if one
composes ${\bf S}$ with block-diagonal canonical transformations, an equivalent DT is obtained, and one recovers the correspondence given by ${\bf\Omega} ={\bf S} {\bf J} {\bf S}^T$ \cite{Bastos,Bernardini13B}.

For a composite system described by the density matrix $\rho$, function of the NC variables $\widehat{z}$, the DT yields the density matrix $\widetilde{\rho} (\widehat{\zeta}) = \rho \left({\bf S}
\widehat{\zeta} \right)$, which is associated with the Wigner function $W \widetilde{\rho} (\zeta)$ 
such that, upon an inversion of the DT, one identifies the NC Wigner function
\cite{Bastos}:
\begin{equation}
W \rho (z) =  \frac{1}{\sqrt{\det \bf\Omega}} W \widetilde{\rho} ({\bf S}^{-1} z) \quad \leftrightarrow \quad
W^{NC}(z)= \frac{1}{\sqrt{\det \bf\Omega}} W({\bf S}^{-1} z),
\label{ggreqs6}
\end{equation}
obtained by {\em turning on} the NC algebra \cite{Bernardini13B,Bernardini13B2,Bastos3}. 
In the simplified scenario where Wigner functions are identified with gaussian functions, if ${\bf \Sigma}$ denotes the covariance matrix of $W \rho$ and ${\bf \widetilde{\Sigma}}$ that of $W \widetilde{\rho}$, then the two are related by:
\begin{equation}
{\bf \Sigma} = {\bf S} {\bf \widetilde{\Sigma}} {\bf S}^T.
\label{ggreqs7}
\end{equation}
A necessary condition for the phase-space Wigner function $W \widetilde{\rho}$ with covariance matrix ${\bf
\widetilde{\Sigma}}$ to be quantum mechanically admissible is that it satisfies the
Robertson-Schr\"odinger uncertainty principle (RSUP) \cite{RSUP1,RSUP2},
\begin{equation}
{\bf \widetilde{\Sigma}} + (i/2) {\bf J} \ge 0,
\label{ggreqs8B}
\end{equation}
that is, in the representation of a $2n \times 2n$ positive matrix in ${C}^{2n}$. For $W \rho$ to be an equally admissible NC Wigner function, from Eqs.~(\ref{ggreqs7}) and (\ref{ggreqs8B}), one concludes that it has to satisfy the NC RSUP,
\begin{equation}
{\bf \Sigma} + (i/2) {\bf \Omega} \ge 0,
\label{ggreqs8}
\end{equation}
which, for NC gaussian states, is also a sufficient condition for defining the quantum behavior.
Here it is worth to mention that a matrix version of the Ozawa uncertainty principle \cite{RSUP1} can also accomodate any NC structure in the phase space \cite{RSUP2}. In particular, it leads to NC corrections to lowest order for two measurement interactions: the backaction evading quadrature amplifier and noiseless quadrature transducers  \cite{RSUP3}. 
\subsubsection*{Entanglement --}
Using similar tools \cite{Bernardini13B}, gaussian separability is identified through a manipulation supported by the introduction of a $2 n\times 2n$ matrix ${\bf \Lambda}= \mathrm{Diag}\left[{\bf I^A},\, {\bf \Lambda^B}\right]$, with ${\bf \Lambda^B}= \mathrm{Diag}\left[{\bf I},\, -{\bf I}\right]$. The transformation
$\zeta \mapsto {\bf \Lambda} \zeta$
amounts to a mirror reflection of $B$ momenta and, through the PPT criterion \cite{Bernardini13B}, if a Wigner function $W \widetilde{\rho} (\zeta)$ is the one of a separable state, then the transformation
\begin{equation}
W \widetilde{\rho} (\zeta) \mapsto W \widetilde{\rho}^{\prime} (\zeta) = W \widetilde{\rho} ({\bf \Lambda} \zeta),
\label{ggreqs11}
\end{equation}
leads to an equivalent Wigner function.
If the state $\widetilde\rho$ is separable then
\begin{equation}
\widetilde{\bf \Sigma}^{\prime} + (i/2){\bf J}\geq 0,
\label{ggreqs12}
\end{equation}
where $\widetilde{\bf \Sigma^{\prime}}$ is the covariance matrix
of $W \widetilde{\rho}^{\prime} (\zeta)$. In terms of the NC variables one has
\begin{equation}
W {\rho} (z) \mapsto W \rho^{\prime} (z) = W \rho ( {\bf D} z),\quad \mbox{with} \quad W \rho^\prime (z) = {1
\over \sqrt{\det {\bf \Omega}}} W \widetilde{\rho}^\prime ({\bf
S}^{-1} z),
\label{ggreqs13}
\end{equation}
where ${\bf
D}= {\bf D}^{-1}= {\bf S}{\bf \Lambda}{\bf S}^{-1} = \mathrm{Diag}[{\bf
I^A},\,{\bf S^B} {\bf \Lambda^B} ({\bf S^B})^{-1}]$, and the covariance matrices ${\bf
\Sigma^{\prime}}$ and ${\bf \widetilde\Sigma^{\prime}}$ of $W
\rho^\prime$ and $W \widetilde\rho^\prime$ are related by ${\bf
\Sigma^{\prime}} = {\bf S}{\bf \widetilde\Sigma^{\prime}} {\bf
S}^T$, which results into a separability condition given exclusively in terms of the NC objects,
\begin{equation}
{\bf \Sigma^{\prime}} + (i/2){\bf \Omega}\geq 0~. \label{ggreqs13A}
\end{equation}
By noticing that ${\bf \Sigma^{\prime}} = {\bf D}{\bf \Sigma} {\bf D}^T$ and 
defining ${\bf \Omega^{\prime}} = {\bf D}^{-1} {\bf\Omega} ( {\bf D}^T)^{-1}$ one also has
\begin{equation}
{\bf \Sigma} + (i/2)   {\bf \Omega^{\prime}}  \ge 0~. \label{ggreqs15}
\end{equation}
with ${\bf \Omega^{\prime}}$ written as
\begin{equation}
{\bf \Omega}^{\prime} = \mathrm{Diag}\left[{\bf
\Omega^A},\,-{\bf \Omega^B}\right],
\label{ggreqs15.1}
\end{equation}
where one has used the definition of ${\bf \Omega}$ and the fact
that ${\bf \Lambda^B}$ is an anti-symplectic transformation, i. e.
${\bf \Lambda^B} {\bf J^B} {\bf \Lambda^B} = - {\bf J^B}$.
By investigating the properties of $W$ and $W^{NC}$, one concludes that a state that is separable in usual QM might be entangled in a NC extension. For gaussian functions, this is a purely kinematical effect, related to the structure of the NC algebra. 


\subsubsection*{Bell Inequalities --}
The above quantum description can be reduced to a bipartite quantum system described in terms of a modes A and B, with collective degrees of freedom $\widehat{z} = (\widehat{z}^A,\,\widehat{z}^B)$, where $\widehat{z}^A =(\widehat{x}, \widehat{p}_x)$ and $\widehat{z}^B=(\widehat{y}, \widehat{p}_y)$, which do help to identify the influence that noncommutativity may have on nonlocal features.
In order to quantify the non-locality of a state through the Clauser, Horne, Shimony and Holt (CHSH) inequality \cite{CHSH}, one must have an explicit definition for the Bell operator. 
The Bell operator is a functional of the Wigner
function which parameterizes a linear combination of four expectation values \cite{CHSH},
\begin{equation}\label{Bell1}
\mathcal{B}={\pi^2\over 4} \left[ W(0,0) +  W(\alpha_1,0) +  W(0,\alpha_2) - W(\alpha_1,\alpha_2)\right]~,
\end{equation}
where $\alpha_{1,2}$ are the phase-space complex amplitudes associated with the operators $\hat{\alpha}_{1}=\hat{x}+i \hat{p_x}$ and $\hat{\alpha}_{2}=\hat{y}+i \hat{p_y}$, that carry two degrees of freedom, $\hat{\alpha}$ and  $\hat{\alpha}^*$ for each mode, $1$ and $2$, and $W(\alpha_1,\alpha_2)$ is the commutative Wigner function of the state calculated in $(\alpha_1,\alpha_2)$.
Local theories admit a description in terms of local hidden variables identified by $\alpha_{1}$ and $\alpha_{2}$ (in Eq.(\ref{Bell1})).
Straightforward calculations \cite{CHSH} allow to show the dichotomic behaviour of the Bell operator corresponds to the boundary value for non-locality given by
$|\mathcal{B}_{min}| > 2$.

In order to implement the NC effects on the Bell functional, one assumes that the Wigner function in Eq.~(\ref{Bell1}) belongs to the set of NC Wigner functions ${\mathcal W}^{NC}$ \cite{Bell}. The
same relation is valid for all observables $\mathcal{O}$ that are completely symmetric in the NC variables\footnote{Let $\widehat \mathcal{O}^{NC}=\mathcal{O}(\widehat z)$ satisfy $\widehat \mathcal{O}^{NC}=(\widehat \mathcal{O}^{NC})_+$, where the subscript $+$ denotes symmetrization with respect to the NC variables, and let $\widehat \mathcal{O}=\mathcal{O}(\widehat\xi)$.}. Then the Weyl symbol $\mathcal{O}^{NC}(z)$ of $\widehat \mathcal{O}^{NC}$ and the Weyl symbol $\mathcal{O}(\xi)$ of $\widehat \mathcal{O}$ are given by the same function, i.e. $\mathcal{O}^{NC}=\mathcal{O}$, and so the average
value functionals of the two observables are the same \cite{Bell}. That is, given a {\it commutative} operator
$\widehat \mathcal{O}=\mathcal{O}(\widehat\xi)$, its NC version is defined by $\widehat \mathcal{O}^{NC}=(\mathcal{O}(\widehat z))_+$, and thus the expectation values of both
operators are given by the same functional of the Wigner function in the NC form.

The time evolution of the NC average values $\langle . \rangle$ is determined from the dynamics of the NC Wigner
function given by
$$
\dot W^{NC}(z;\,t)= [H^{NC}(z), W^{NC}(z;\,t)]_{NC} \, ,
$$
where $H^{NC}(z) = \tilde H({\bf S}^{-1} z)$, $\tilde H(\xi)$ is the Weyl symbol of $H({\bf S} \widehat\xi)$, and the (commutative and the NC) Moyal brackets shall determine two different dynamics for the Wigner
functions.
To quantifiy the impact of noncommutativity on the dynamics of the Bell function, one assumes that at the initial time
the commutative and the NC configurations are the same. 
Therefore, at $t=0$ one has $W(x,\,p;\,t=0)=W^{NC}(x,\,p;\,t=0)$, and so $\mathcal {B}^{NC}(t=0)\equiv
\mathcal{B}(W^{NC}(x,\,p;\,t=0))=\mathcal{B}(W(x,\,p;\,t=0))\equiv \mathcal{B}^C(t=0)$.
In this case, the measurement of
expectation values of the fundamental variables is not influenced by the {\it a priori} knowledge of the NC QM structure.
The only restriction is that the (NC) Wigner function yields the observed averaged values, and the time evolution of commutative and NC Wigner functions can be obtained in order to compute
$\mathcal{B}^{NC}(t)=\mathcal{B}(W^{NC}(x,\,p;\,t))$, which can be compared to $\mathcal{B}^C(t)=\mathcal{B}(W(x,\,p;\,t))$.
However, due to the non-trivial nature of the Bell operator, the inclusion of NC effects into its properties deserves further investigations. The results obtained in Ref. \cite{Bell} indicate that the non-locality encountered in the NC case is qualitatively similar to the one of the usual QM.

\subsubsection*{Additional Issues and Outlook --}
\par
Extending the known results of QM to the phase space NC QM framework requires investigating various quantum mechanical phenomena in order to fully understand the implications of the algebra deformation. For instance, examining whether gauge invariance or the equivalence principle hold under the deformation of the Heisenberg-Weyl algebra is particularly relevant. Regarding the former, when considering a particle in a magnetic field, it has been shown that in order to preserve gauge invariance, one must impose the vanishing of the configuration space noncommutative parameter, $\theta$ \cite{Leal}. This follows from the fact that the Hamiltonian for the problem,
\begin{equation}
\hat{H}=\frac{1}{2m}\left[\boldsymbol{p}-q\boldsymbol{A}(\boldsymbol{x})\right]^2,
\end{equation}
under a DT gives rise to terms that are not gauge invariant. 
As for the equivalence principle, it is shown that, when a gravitational field is considered in the single particle Hamiltonian, a suitable change to an accelerated frame of reference, $x\rightarrow x'=x+\sigma(t)$, is able to absorb this potential with only a phase difference on the wave function, that consequently is negligible for physical observables. Thus, the weak equivalence principle is recovered in NCQM \cite{Leal}.
 
Other known results of standard QM should be addressed in the context of NCQM. Interesting cases involve issues related with the measurement problem \cite{Ballentine}. This will be pursued in a future work.

\vspace{.3 cm}
{\em Acknowledgments} --  The work of OB is partially supported by the COST action MP1405. The work of AEB is supported by the Brazilian agency FAPESP (grant 17/02294-2). The work of PL is supported by FCT (Portuguese Funda\c{c}\~ao para Ci\^encia e Tecnologia) grant PD/BD/135005/2017.

\smallskip
\end{document}